\documentclass[twocolumn,showpacs,preprintnumbers,amsmath,amssymb,prb]{revtex4}

\usepackage{graphicx}% Include figure files
\usepackage{dcolumn}% Align table columns on decimal point
\usepackage{bm}% bold math

\newcommand{\up}{\uparrow}
\newcommand{\dn}{\downarrow}

\begin{document}

% \preprint{APS/123-QED}

\title{Spin-triplet $s$-wave local pairing induced by Hund's rule coupling}

\author{J. E. Han}
\affiliation{Department of Physics, The Pennsylvania State University, 
University Park, PA 16802-6300, USA\\
Department of Physics, State University of New York at Buffalo, 
Buffalo, NY 14260, USA}

\date{\today}

\begin{abstract}
We show within the dynamical mean field theory that 
local multiplet interactions such as 
Hund's rule coupling produce local pairing
superconductivity in the strongly correlated regime.
Spin-triplet superconductivity driven by the Hund's rule coupling
emerges from the pairing mediated by local fluctuations
in pair exchange. In contrast to the conventional spin-triplet theories,
the local orbital degrees of freedom has the
anti-symmetric part of the exchange symmetry, 
leaving the spatial part as fully gapped and symmetric $s$-wave.
\end{abstract}


\maketitle

\section{Introduction}

For the last two decades, discovery of unconventional superconductors have 
fueled intensive research for novel pairing mechanisms. Of particular
interest have been the cuprate~\cite{damascelli} and 
heavy-fermion~\cite{steglich} systems where
the very existence of superconductivity is quite surprising 
due to the strong Coulomb repulsion
within the framework of the conventional Eliashberg-Migdal 
theory~\cite{schrieffer}.
Recently discovered class of 
superconductors close to the ferromagnetic phase such as 
Sr$_2$RuO$_4$~\cite{mckenzie}, UGe$_2$~\cite{saxena} demand better 
understanding between symmetry and superconductivity which goes beyond 
the $s$-wave BCS theory.

One of the most puzzling aspects of the superconductivity in the
strongly correlated regime is the role of the repulsive Coulomb 
interaction. Common belief is that the pairing mechanism is driven
or assisted by the Coulomb repulsion to an attractive 
interaction between electrons which results in spatial or temporal 
structures in the superconducting wavefunction to avoid direct Coulomb
repulsion. 
Even some conventional $s$-wave phonon-mediated superconductors 
such as alkali-doped fullerenes have also shown extraordinarily high 
transition temperatures~\cite{hebard} despite the strong Coulomb 
repulsion near the Mott transition~\cite{lof,murphy}.

Even before the onset of superconductivity,
direct Coulomb interaction is heavily renormalized away 
by electrons moving in a correlated manner to form a narrow 
resonance near chemical potential.
The superconductivity in such regime should be understood in terms of those
correlated electronic basis, typically called quasi-particles (QP).
The problem of mutual renormalization effects between electron-electron 
and electron-phonon (el-ph) has received a great deal of interest resulting
in different interpretations, ranging from the conventional
Coulomb pseudopotential~\cite{schrieffer,pseudoU}, field-theoretic 
approaches~\cite{tesanovic,newns} to numerical analysis~\cite{freericks}.
Recently, it has been pointed out~\cite{han_sc} that the symmetry of the
el-ph interaction critically influences the pairing interaction where
the two interactions are so interwoven that the effective pairing 
interaction cannot be reduced into a form similar to the McMillan
formula~\cite{mcmillan}.

We show here on the basis of generic hamiltonians that local orbital
correlations are crucial for the superconductivity and, in particular, 
predict the Hund's rule (HR) induced superconductivity.
We discuss the local pairing as mediated by self-generated fluctuations
within the on-site interactions and show that the Coulomb interaction
and multiplet interaction work in fundamentally different way
than described in perturbative theories.
The HR superconductivity has the spin-triplet pairing state. In 
contrast to the well-known non-$s$-wave triplet pairing~\cite{balian},
the anti-symmetric part of the exchange symmetry of the pairing 
wavefunction is taken up by the local orbital variables, 
leaving the spatial part
as the symmetric $s$-wave. We will discuss possible consequences
of this new pairing structure.

This paper is organized as follows. In section~\ref{sec:models},
we define the problem in models of multiplet interactions
for Hund's rule and Jahn-Teller el-ph couplings. In the following
section~\ref{sec:pairing}, the pairing mechanism is explained in
a perturbation theory of a simplified dynamical mean field theory (DMFT).
Results from full DMFT calculations are presented in the 
section~\ref{sec:results} and are discussed in detail. 
In Appendices~\ref{app1}and \ref{app2}, we describe the calculations 
for the local energy levels and perturbation theory. Finally, a new
Hubbard-Stratonovich decoupling scheme for the Hund's rule coupling
which bypasses the sign-problem is introduced in Appendix~\ref{app:hs}.

\section{models}
\label{sec:models}

We consider cases of doubly degenerate band electrons at half-filling
which are locally coupled via Hund's rule (HR) and Jahn-Teller (JT) 
phonons.
We model the multi-band electronic system with the on-site
Coulomb repulsion of strength $U$ as
\begin{eqnarray}
H & = & -\sum_{ijm\sigma}t_{ij}d^\dagger_{im\sigma}d_{jm\sigma}
        -\mu\sum_i N_i \nonumber \\
  & &   +{U\over 2}\sum_iN_i(N_i-1)+\sum_iH_{\rm mult}(i),
\label{eq:hfull}
\end{eqnarray}
where $d^\dagger_{im\sigma}\,(d_{im\sigma})$ is the electron creation
(annihilation) operator acting on site $i$, orbital $m\,(=1,2)$, spin
$\sigma$,  $\mu$ is the chemical potential
and $N_i=\sum_{m\sigma}n_{im\sigma}$. 
The hopping integral $t_{ij}$ is chosen to give a semi-elliptical
density of states with the half-bandwidth $W/2=1$~\cite{DMFT}.
The local multiplet interaction is taken as
$H_{\rm mult}=H_{\rm HR}$ or $H_{\rm JT}$~\cite{han_sc},
\begin{eqnarray}
H_{\rm HR} & = & K\left[\sum_{m}n_{m\uparrow}n_{m\downarrow}
                 \!-\!\!\!\sum_{m\neq m'}\!\!n_{m\uparrow}n_{m'\downarrow}
                 \!-2\!\!\!\!\sum_{\sigma,m<m'}\!\!\!\!n_{m\sigma}
		 n_{m'\sigma}\right] \nonumber \\
           & + & K\!\!\sum_{m\neq m'}\!\!\left(
                 d^\dagger_{m\uparrow}d^\dagger_{m\downarrow}
                 d_{m'\downarrow}d_{m'\uparrow}
                +d^\dagger_{m\uparrow}d^\dagger_{m'\downarrow}
                 d_{m\downarrow}d_{m'\uparrow}\right), \label{eq:hhr}\\
H_{\rm JT} & = & {1\over 2}\sum_\nu\left(\dot{\varphi_{\nu}}^2
                 +\omega_{\rm ph}^2\varphi_{\nu}^2\right) \nonumber \\
           &+&   \!\!\!{g\over\sqrt{\omega_{\rm ph}}}
                 \!\sum_{\sigma}\!\!\left[\varphi_{1}(n_{1\sigma}
                 -n_{2\sigma})+\varphi_{2}(d^\dagger_{1\sigma}d_{2\sigma}
                 +{\rm h.c.})\right], \label{eq:hjt}
\end{eqnarray}
where the site index $i$ has been omitted for brevity.
$K$ is the HR coupling constant and $\varphi_\nu$ $(\nu=1,2)$ 
are the phonon fields with the bare frequency 
$\omega_{\rm ph}$, and $g$ the el-ph coupling constant.
In the anti-adiabatic limit of phonons ($\omega_{\rm ph}\rightarrow
\infty$)~\cite{tosatti}, $H_{\rm JT}$ maps
to the same form as $H_{\rm HR}$ but with a negative (fictitious)
$K_{\rm JT} =-2\Delta$ ($\Delta\equiv g^2/\omega_{\rm ph}$).
Therefore, the el-ph superconductivity is suppressed as
the physical HR coupling is turned up. 
We will show that, as $K$ increases further,
superconductivity re-emerges in a spin-triplet channel.

\section{Local pairing mechanism}
\label{sec:pairing}

We describe main ideas behind the local pairing before
presenting results from the full dynamical mean field theory.
In the strongly correlated regime, QP states within the energy band
$[-zW,zW]$ ($z$ the QP renormalization factor, $W$ non-interacting
bandwidth) are usually responsible for low-energy manybody phenomena.
The effect of $U$ is implicit in the renormalization factor $z$ and
the truncated high energy states.
In such limit, we simplify the DMFT by singling out the QP band and ignore
incoherent high energy excitations as
\begin{equation}
  H_{\rm eff} = H_{\rm int} \!+ 
		\!\!\sum_{m{\bf k}\sigma}\!\!
 		\left[\epsilon_{q{\bf k}}n_{m{\bf k}\sigma}
               \!+\!{t_q\over \sqrt{N_s}}(
               c^\dagger_{m{\bf k}\sigma}d_{m\sigma}\!+h.c.)\right],
\end{equation}
where $c^\dagger_{m{\bf k}\sigma}\,(c_{m{\bf k}\sigma})$ is creation 
(annihilation) of a QP state, $\epsilon^q_{\bf k}$
is QP energy (absorbing chemical potential), $t_q\,(\sim zW)$ is 
hopping integral on and off the impurity and $N_s$ is the number of sites.
Contributions from the high energy states are correctly considered
in the full DMFT calculations.
$H_{\rm int}$ is the interacting part of the original hamiltonian.
Note that the interaction only 
acts on the impurity site. For small QP bandwidths with $zW$ smaller
than other energy scales, we further simplify the above effective 
hamiltonian to a two-site model,
\begin{equation}
  H_{\rm int} = H_{\rm mult} + \epsilon_q N_c+t_q \sum_{m\sigma}(c^\dagger_{m\sigma}d_{m\sigma}+h.c.),
\label{eq:heff}
\end{equation}
with $c^\dagger_{m\sigma}=\sqrt{N_s^{-1}}\sum_{\bf k}
c^\dagger_{m{\bf k}\sigma}$,
$N_c=\sum_{m{\bf k}\sigma}n_{m{\bf k}\sigma}$ and $\epsilon_q$ 
the QP energy ($\epsilon_q\approx 0$), where the narrow QP band is
represented by a single energy level. $H_{\rm eff}$
has the similar construction as the exact diagonalization implementation
of the DMFT, where the bath spectrum is sampled with a finite number of energy 
levels~\cite{DMFT}. Representing the QP states with several discrete levels
over a small energy range made little change. 
This procedure is expected to be
valid as long as the electrons slow down by the strong Coulomb 
interaction so that the local picture is meaningful and there is a QP
reservoir which supplies particles at an infinitesimal energy cost.

We solve the two-site problem Eq. (\ref{eq:heff}) using the exact
diagonalization method. We analyze the solution as a perturbation expansion 
in $t_q$~\cite{metzner}. We estimate the pairing 
correlation by computing the energy gain $E_p$ for creating two extra
electrons on a half-filled impurity site,
\begin{equation}
 E_p = [E(N_d\!+\!2)-E(N_d)]-2[E(N_d\!+\!1)-E(N_d)],
\label{eq:corr}
\end{equation}
where $E(n)$ is the ground state energy with $n$ total electrons
and $N_d$ is the orbital degeneracy ($N_d=2$).
It is crucial to have multiple electrons per site for the local multiplets
to induce the local pairing.
As shown in Ref.~\cite{han_sc}, the local pairing
has a strong filling dependency.

\begin{figure}[bt]
\rotatebox{0}{\resizebox{3.0in}{!}{
\includegraphics{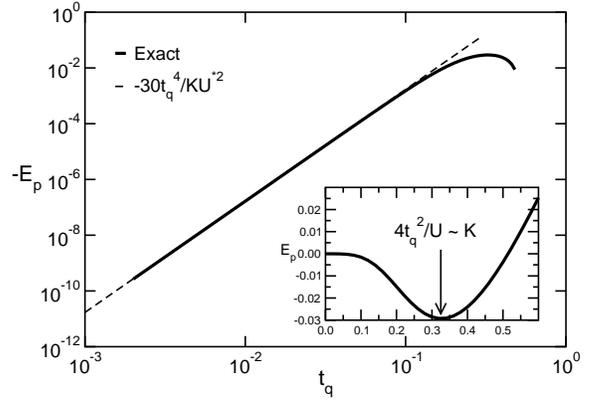}}}
\caption{
Pair correlation energy $E_p$ for two-site model Eq.~(\ref{eq:heff})
with the Hund's rule coupling. $E_p$ is attractive for a wide
range of hopping integral to the quasi-particle bath, $t_q$. 
Coupling to quasi-particle bath (thick line) produced a good
agreement with the perturbation theory, $-30t_q^4/KU^{*2}$
($U^{*}=U+2K$).
The optimal interaction is at $4t_q^2/U\sim K$ for the maximum
pair attraction $E_p\propto -K$.
}
\label{fig:pair}\end{figure}

\begin{figure}[bt]
\rotatebox{0}{\resizebox{3.0in}{!}{
\includegraphics{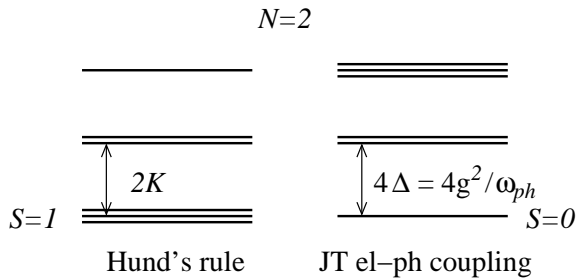}}}
\caption{Single site energy level scheme for two on-site electrons
($N_d=2$) with Hund's rule and Jahn-Teller couplings. The degeneracy of 6
is lifted by the multiplet interaction $H_{\rm mult}$.
The ground states are spin-triplet ($S=1$) and spin-singlet ($S=0$)
for Hund's rule. The Jahn-Teller multiplets have inverted level structure
of the Hund's rule coupling.
}
\label{fig:scheme}\end{figure}

\begin{figure}[bt]
\rotatebox{0}{\resizebox{3.0in}{!}{
\includegraphics{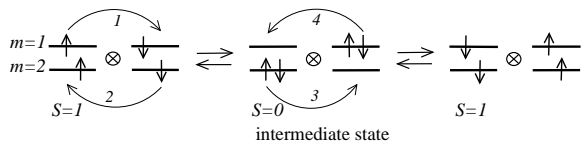}}}
\caption{One of the contributions in $E(N_d+2)$ for
the pair correlation energy $E_p$
in Eq.~(\ref{eq:epair}). Two extra electrons created on the interacting
site dissolve into the quasi-particle bath to avoid Coulomb 
repulsion, which produces the ground state configurations shown on the
left and right panel. 
The two unperturbed degenerate ground states couple
via exchange of the spin-triplet electron pairs in the 4-th order perturbation 
of $t_q$ (from the left configuration to the right).
In the intermediate state, the spins on the interacting site are in
a locally excited state (See Fig.~\ref{fig:scheme}).
}
\label{fig:dia}\end{figure}

Numerical results shown for the HR problem are
with parameters of $U=4$ and $K=0.1$.
For small $t_q$ in a HR system, $E_p$ is negative
(thick line in Fig.~\ref{fig:pair}), {\em i.e.,} it is energetically
favorable to create mutually interacting quasi-particles than to create
two independent quasi-particles.
The impurity ground states are always predominantly of filling $N_d(=2)$ 
with the total spin $S=1$ since any extra 
electrons created on the impurity are repelled by the Coulomb interaction
and are immediately dissolved into the QP bath.
Therefore we first examine the atomic limit ($t_q=0$). Without $H_{\rm mult}$
the degeneracy of two-electron configurations is 6. The HR coupling 
lifts the degeneracy with the ground states of spin-triplet 
$(S=1)$ on the impurity site with
\begin{equation}
|T_m)=\left\{d^\dagger_{1\uparrow}d^\dagger_{2\uparrow},
{1\over\sqrt{2}}\left(d^\dagger_{1\uparrow}
        d^\dagger_{2\downarrow}-d^\dagger_{2\uparrow}
        d^\dagger_{1\downarrow}\right),
d^\dagger_{1\downarrow}d^\dagger_{2\downarrow}\right\}|vac),
\label{eq:jtgr}
\end{equation}
with $m=+1,0,-1$, respectively (see FIG.~\ref{fig:scheme} and 
Appendix~\ref{app1}). 
The 1st excitation energy is $2K$ for the HR coupling. 
The local energetics is described in Appendix~\ref{app1}.
Upon adding two electrons to the interacting site, the extra charges
are immediately pushed to the QP reservoir to avoid the Coulomb repulsion.
Through intersite hopping, the ground state triplet degeneracy is lifted.
The resulting leading order ground state wavefunction has the most 
symmetric form ($U\gg K \gg t_q>0$),
\begin{equation}
  |\psi_0^{N_d+2}\rangle = {1\over\sqrt{3}}\left\{
        |T_1)\otimes |t_{\bar{1}})+|T_0)\otimes |t_0)
        +|T_{\bar{1}})\otimes |t_1)\right\},
\label{eq:gr}
\end{equation}
where the second ket-vectors for the QP states in the direct products
have similar definitions as Eq.~(\ref{eq:jtgr}).
We analyze the perturbed ground state wavefunction and its energy
in terms of the perturbation theory developed in Appendix~\ref{app2}.
The first order perturbed wavefunction becomes from Eq.~(\ref{app:psi1}),
\begin{eqnarray}
|\psi_1^{N_d+2}\rangle & = & {\sqrt3t_q\over U^*}\left(
d^\dagger_{1\up}d^\dagger_{2\up}d^\dagger_{1\dn}c^\dagger_{2\dn}-
d^\dagger_{2\up}c^\dagger_{1\up}c^\dagger_{1\dn}c^\dagger_{2\dn}\right. 
\label{eq:psi1}\\
&+ & \left. d^\dagger_{1\up}c^\dagger_{2\up}d^\dagger_{1\dn}d^\dagger_{2\dn}-
c^\dagger_{1\up}c^\dagger_{2\up}d^\dagger_{2\dn}c^\dagger_{1\dn}-
(1\leftrightarrow 2)\right)|vac\rangle, \nonumber
\end{eqnarray}
with the ionization energy $U^*/2=U/2+K$.
This intermediate state has an ionized impurity at the excited energy
$U^*$.
With Eq.~(\ref{app:e2}) and $E_0-H_0=-U^*/2$, one gets the second order
contribution to the energy
\begin{equation}
  E_2=\left({\sqrt3 t_q\over U^*}\right)^2 (-U^*/2)\cdot 8 = -{12t_q^2\over
  U^*},
\end{equation}
where the figure 8 is from the number of terms in Eq.~(\ref{eq:psi1}).
The second order correction to the wavefunction from Eq.~(\ref{app:psi2})
becomes
\begin{equation}
|\psi_2^{N_d+2}\rangle = {2\sqrt3 t_q^2\over KU^*}\left[
|S_a)\otimes|s_a)+|S_b)\otimes|s_b)-{1\over 2}|S_c)\otimes|s_c)\right],
\label{eq:psi2}
\end{equation}
where $|S_a),\,|S_b)$ and $|S_c)$ are spin-singlet excited states on the
interacting site as defined in Appendix~\ref{app1}. The QP configurations
$|s_a),\,|s_b),\,|s_c)$ are defined with the same symmetry as the impurity
states. 

One of the processes in the second order contribution is depicted in 
Fig.~\ref{fig:dia}. Typical (unperturbed) basis states in the ground 
state are spin-triplets as shown on the left and right side. 
The two states couple via the electron hopping where a second-order
intermediate state
is shown in the middle. After two hoppings ({\em eg.} the processes 1 and 2 
from the left configuration), the resulting intermediate state 
becomes a charge-neutral spin-singlet ($S=0$), which correspond
to the term $|S_a)\otimes|s_a)$ in Eq.~(\ref{eq:psi2}) and a
local transition between the multiplets in Fig.~\ref{fig:scheme}.
Other charge-excited intermediate states have small contributions
to the transition amplitude (of order $K/U$ smaller).

From Eq.~(\ref{app:e4}), the forth order correction to the energy becomes
\begin{eqnarray}
 E_4 & = & \left({2\sqrt3 t_q^2\over KU^*}\right)^2\left[-2K-2K-4K
     \left({1\over 2}\right)^2\right] \nonumber \\
    & & -{12t_q^2\over U^*}8\left({\sqrt3 t_q\over U^*}\right)^2
     \approx  -{60\,t_q^4\over K U^{*2}},
\end{eqnarray}
for $K\ll U^*$.
The ground state energy $E(N_d+2)$ becomes
\begin{equation}
  E(N_d+2) = U-{12\,t_q^2\over U^*}-{60\,t_q^4\over K U^{*2}}.
\label{eq:en2}
\end{equation}
The first term is the (unperturbed) atomic-limit.
The last term contributes to the pairing interaction.
The denominator in Eq.~(\ref{eq:en2}) comes from 
two ionized states (with charging energy $U^*$ after the hoppings 1 and 
3 in Fig.~\ref{fig:dia} for example) 
and one HR-excited state (with excitation energy $\sim 2K$ after the
hopping 2). 

Ground state for one extra (spin-up) electron can be obtained similarly.
For hopping integrals smaller than other energy scales, the interacting site 
is predominantly of filling 2 and of spin-triplet. Therefore the
ground state consists of $|T_0)\otimes |c_{m\uparrow})$ and $|T_1)\otimes 
|c_{m'\downarrow})$ ($|c_{m\sigma})\equiv c^\dagger_{m\sigma}|vac)$). 
Due to the hopping interaction, states of total spins $S_z=\pm 3/2$
do not appear in the ground state.
An explicit calculation gives the unperturbed ground state,
\begin{equation}
|\psi_0^{N_d+1}\rangle={1\over\sqrt{3}}\left\{|T_0)\otimes 
|c_{1\uparrow})+\sqrt{2}|T_1)\otimes |c_{2\downarrow})\right\}.
\end{equation}
The larger coefficient for $|T_1)\otimes |c_{2\downarrow})$ reflects
more paths for hopping transitions for this configuration.
Similar calculations as those with $N_d+2$ electrons give
\begin{eqnarray}
|\psi_1^{N_d+1}\rangle\!\! & = & \!\!{2 t_q\over \sqrt6 U^*}\left\{
   3d^\dagger_{1\up}d^\dagger_{2\up}d^\dagger_{2\dn}
  +3d^\dagger_{1\up}c^\dagger_{2\up}c^\dagger_{2\dn} \right. \nonumber \\
  & & \left.-2d^\dagger_{2\up}c^\dagger_{1\up}c^\dagger_{2\dn}
  -d^\dagger_{2\up}c^\dagger_{2\up}c^\dagger_{1\dn}
  +c^\dagger_{1\up}c^\dagger_{2\up}d^\dagger_{2\dn}\right\}|vac\rangle \\
|\psi_2^{N_d+1}\rangle\!\! & = & \!\!{\sqrt3 t_q^2\over KU^*}\left[
   |S_a)\otimes|c_{2\up})+|S_b)\otimes|c_{1\up})-{
   |S_c)\otimes|c_{1\up})\over 2}\right] \\
E(N_d+1) & = & U -{8t_q^2\over U^*}-{15t_q^4\over K U^{*2}}. 
\end{eqnarray}
The state without extra electrons results in no energy gain in the 4-th order,
\begin{eqnarray}
|\psi^{N_d}\rangle & = & \left(1+{2t_q\over U^*}\sum_{n\sigma}c^\dagger_{n\sigma}d_{n\sigma}\right)|T_m)\otimes|vac) \\
E(N_d) & = & U-{4t_q^2\over U^*}.
\end{eqnarray}
Finally, the pairing correlation Eq.~(\ref{eq:corr}) becomes
\begin{equation}
 E_p=-30t_q^4/KU^{*2},
\label{eq:epair}
\end{equation} 
which is in excellent agreement
with the exact diagonalization results in the $t_q\ll K$ limit.
It is interesting to note that $E_p$ becomes smaller for stronger
$K$ due to the higher multiplet excitation energy.
The exact diagonalization results show maximum attractive pair correlation
for $4t_q^2/U\sim K$, where the effective hopping exchange energy is 
comparable to the internal excitation energy. 
Upon substitution for $t_q$ in Eq.~(\ref{eq:epair}), this relation 
gives $E_p\propto -K$ at the optimal pairing.

One can also interpret the pairing as arising from pair
exchange. In Fig.~\ref{fig:dia}, the 
spin-up ($S_z=1$) pair on the interacting site hops out via processes 
1 and 3, while the spin-down pair hops back in via 2 and 4. 
The charge neutral intermediate state forces the temporal overlap
between the hoppings of the pairs, producing the attractive interaction
when viewed as a second order perturbation in terms of the pair-basis.
The same general reasoning should apply to other types of local interactions
in the strong-$U$ limit.

One can carry out the same perturbation analysis for the Jahn-Teller
interaction Eq.~(\ref{eq:hjt}), where the pairing energy gain $E_p$
is always negative. $E_p=-7t_q^4/\Delta {U^*}^2$ for $U\gg\Delta\gg t_q$,
which retains the similar form as Eq.~(\ref{eq:epair}) for the HR 
coupling. Leading order wavefunctions and their perturbed
energies are written down in Appendix~\ref{app2}.
On the other hand, for a pure Hubbard model without the 
multiplet terms ($H_{\rm mult}=0$), $E_p=344 t_q^4/U^3>0$
suggesting no superconductivity, which is also
supported by full DMFT calculations~\cite{DMFT}. With $H_{\rm mult}=0$,
the electrons only interact with charge fluctuations making
the local internal fluctuations irrelevant. 
The above perturbation results are summarized in Table 1.

\begin{table}[h]
\begin{tabular}{c|ccc} \hline\hline
 & $K=\Delta=0$ & $K > 0$ & $\Delta > 0$\\ \hline\hline
$U=0$ & 0 & \,\,- & $\,\,-2\Delta$ \\ \hline
$t_q^2/U\ll\Delta, K$\ \ \ \  & \ \ $344t_q^4/U^3$\ \  &
\ \ $\,\,-30t_q^4/KU^2$ & $\,\,-7t_q^4/\Delta U^2$ \\ \hline\hline
\end{tabular}
\caption{\small The pair correlation energy for an effective 2-site
2-band model with Hund's rule ($K>0$) and the Jahn-Teller el-ph 
($\Delta>0$) coupling.
For large $U$, both of the couplings produce attractive pairing energy.
($\Delta=g^2/\omega_{\rm ph}$.)}
\end{table}

As the pairing correlation energy Eq.~(\ref{eq:epair}) and Table 1 suggest,
the Coulomb interaction and the multiplet interaction (including the
JT el-ph interaction) cannot be separated into a form similar to
$\lambda-\mu^*$ as in the McMillan formula~\cite{mcmillan}, where
the Coulomb interaction represented as the Coulomb pseudopotential
$\mu^*$ interact additively with the multiplet interaction 
({\em eg.} el-ph coupling) represented by the mass-renormalization 
factor $\lambda$.

\section{Pairing instability: DMFT results}
\label{sec:results}

Although the 2-site model is suggestive of the nature of the pairing,
we show the existence of superconductivity
from a lattice calculation. 
The lattice hamiltonian Eq. (\ref{eq:hfull}) is solved within the DMFT 
which approximates one-particle self-energies and vertex corrections 
as momentum-independent and maps a lattice model to an effective 
quantum impurity problem.
We solve the impurity problem using quantum Monte Carlo (QMC) 
technique without making any physical 
approximations other than discretizing the imaginary time.
It has been known that the QMC technique often suffers from the 
sign-problem~\cite{signproblem}. DMFT-impurity models
with the Hubbard and electron-phonon couplings
have been known not to cause any serious sign problem for most of 
the physical parameter space~\cite{han_sc}. 
However, inclusion of the Hund's rule coupling terms
in Eq.~(\ref{eq:hhr}) produced severe sign-problems~\cite{han_DMFT} 
when each of the
interaction terms are decoupled by the discrete Hubbard-Stratonovich
transformation~\cite{hirsch}, despite that the decoupled HR terms 
resemble the Jahn-Teller coupling. 
To overcome such problem, a new scheme
of mixing discrete~\cite{hirsch} and continuous~\cite{bss} 
Hubbard-Stratonovich transformations
has been used. This removed the problem with the average signs larger
than 0.9 for most of the runs. Details of the procedure is given in
Appendix~\ref{app:hs}. It is speculated that the simulation is more
stabilized by the built-in symmetry of orbitals in the new
transformation in contrast to the poorly preserved orbital symmetry 
in incomplete QMC samplings of the old scheme.

\begin{figure}[bt]
\rotatebox{0}{\resizebox{3.3in}{!}{
\includegraphics{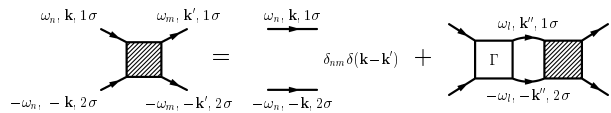}}}
\caption{Expansion of spin-triplet pairing susceptibility $\chi$. 
The local index $m\sigma$ ($m=1,2$) is for orbital and spin labels.
}
\label{fig:diagram}\end{figure}

The superconducting instability is probed
by computing  uniform pair susceptibility $\chi$. 
$\chi$ for the spin-triplet channel can be expressed in a Bethe-Salpeter 
equation as
\begin{eqnarray}
  \chi & = & \chi_0 + \chi_0\Gamma\chi_0 + \chi_0\Gamma\chi_0\Gamma\chi_0 +\cdots \nonumber \\
       & = & \sqrt{\chi_0}(I-\sqrt{\chi_0}\,\Gamma\sqrt{\chi_0})^{-1}\sqrt{\chi_0},
\label{eq:bethe}
\end{eqnarray}
where we have used the symmetrized form~\cite{scalapino}.
$\chi_0$ is the propagator of two electrons independently
propagating with zero net momentum, and $\Gamma$ is the 
effective pairing interaction. (See Fig.~\ref{fig:diagram}.)
Eq.~(\ref{eq:bethe}) is a shorthand for matrices with indices
over the Matsubara frequency $\omega_n$.
For the static uniform susceptibility, we only need in-coming
electrons of net zero momenta and net zero frequency, as in
Fig.~\ref{fig:diagram}. Due to the momentum and frequency
conservation at each internal vertices of Feynman diagrams,
any internal pair of legs in the ladder diagrams in 
Fig.~\ref{fig:diagram} also have net zero momenta and frequencies.
The uniform (uncorrelated) 2-particle propagator
$\chi_0$ can be computed with the 1-particle
Green function as $\chi_0(\omega_n)=\sum_{\bf k}G({\bf k},\omega_n)
G({\bf -k},-\omega_n)$. The ${\bf k}$-summation is performed within
$\chi_0$ due to the locality of the DMFT.
$\Gamma$ is approximated by a local quantity $\Gamma^{\rm loc}$
obtained from a local relation similar to Eq.~(\ref{eq:bethe}),
\begin{equation}
\Gamma^{\rm loc}=[\chi_0^{\rm loc}]^{-1}-[\chi^{\rm loc}]^{-1}.
\label{eq:gamma}
\end{equation}
$\chi_0^{\rm loc}$ is the local (uncorrelated) 2-particle propagator,
$\chi_0^{\rm loc}(\omega_n)=G^{\rm loc}(\omega_n)
G^{\rm loc}(-\omega_n)$,
and the local triplet pair susceptibility $\chi^{\rm loc}$ is 
directly computed from QMC as
\begin{equation}\label{eq:chiT}
\chi^{\rm loc}(\tau_1,\tau_2,\tau_3,\tau_4)=  \!
\langle T\lbrack d_{1\sigma}^{\dagger}(\tau_1)
d_{2\sigma}^{\dagger}(\tau_2) d_{2\sigma}(\tau_3)
d_{1\sigma}(\tau_4)\rbrack \rangle.
\end{equation}
Corresponding definition of $\chi^{\rm loc}$ for the singlet channel
can be found in Ref.~\cite{han_sc}.
Since $\chi^{\rm loc}$ are directly
obtained from the QMC measurements and $\chi_0,\,\chi_0^{\rm loc}$ 
are easily computed from the 1-particle self-energy,
we derive the effective
interaction kernel $\Gamma\,(\equiv \Gamma^{\rm loc})$ 
and the uniform pair susceptibility
$\chi$ from Eqs.~(\ref{eq:bethe},\ref{eq:gamma}).

The transition temperatures are determined by when the maximum eigenvalue 
of the superconducting kernel $\sqrt{\chi_0}\,\Gamma\sqrt{\chi_0}$ 
in Eq.~(\ref{eq:bethe})
approaches 1. For such condition, the effective interaction must be
attractive ($\Gamma>0$). The other requirement is the
finite weight of quasi-particles at the chemical potential
as in $\chi_0(\omega)\approx \pi z 
N(0)/\omega$, with $N(0)$ non-interacting density of states at the
chemical potential. As the system becomes strongly correlated, $\chi_0$
becomes smaller as $z\rightarrow 0$ while the effective local correlation
tends to get larger as electrons are more localized.
The resulting $T_c$ is determined by the balance between
$\chi_0$ and $\Gamma$.

\begin{figure}[bt]
\rotatebox{0}{\resizebox{3.0in}{!}{
\includegraphics{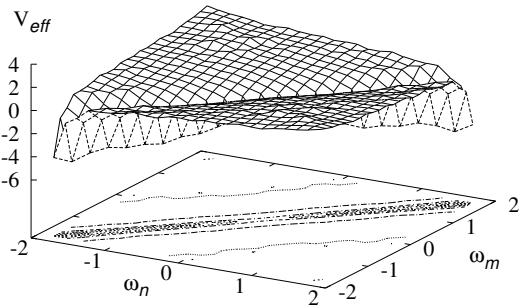}}}
\caption{The effective interaction $V_{\rm eff}(\omega_n,\omega_m)\,(\equiv
-\Gamma(\omega_n,\omega_m)/T)$. The attractive interaction is
for small energy transfer ($\omega_n\approx\omega_m$). The interaction
at higher frequencies is repulsive.
}
\label{fig:veff}\end{figure}

The attractive interaction $V_{\rm eff}(\omega_n,\omega_m)\,(\equiv -(1/T)
\Gamma(\omega_n,\omega_m))$ is shown in Fig.~\ref{fig:veff} for
$U/W=0.6$ and $K/W=0.1$. It is quite interesting that 
$V_{\rm eff}(\omega_n,\omega_m)$ resembles that for the electron-phonon coupling
in the Eliashberg theory~\cite{schrieffer} where $V_{\rm eff}(\omega_n,\omega_m)
=g^2D(\omega_n-\omega_m)$ with the phonon propagator $D(\omega_n)$.
The attractive part is formed in a uniform attractive valley along 
the small energy transfer region $\omega_n\approx\omega_m$.
For pure Hubbard interactions, the effective interaction
$V_{\rm eff}(\omega_n,\omega_m)$ shows a very different profile with 
high energy structures~\cite{jarrell} away from the low energy scattering 
$\omega_n=\omega_m$. The attractive interaction along
the $\omega_n=\omega_m$ line clearly demonstrates that the HR-induced
local pairing is mediated by the effective low-energy coupling medium which
has been self-generated by the multiplet interaction.
Since the local spin excitation energy is $2K$ 
(see Fig.~\ref{fig:scheme}) and the electron-spin coupling
constant is of order $K$, a naive estimate
for the effective interaction gives $\sim K^2(2/2K)\sim K(=0.2)$,
as in the electron-phonon theory. The numerically obtained
$V_{\rm eff}$ at $2-4$ is about an order of magnitude larger than
the simple estimate. This suggests that the renormalization
effect is very strong and is consistent with the above
observation that the suppressed hopping due to the Coulomb
interaction enhances the local interaction~\cite{veff}. 
A thorough examination of the renormalization effects
is necessary to fully understand these numerical
results. Diagrammatic methods will prove particularly
useful when backed up by the QMC and the exact diagonalization 
methods since the basic mechanism of the local pairing is given 
as in Section III.

\begin{figure}[bt]
\rotatebox{0}{\resizebox{3.0in}{!}{
\includegraphics{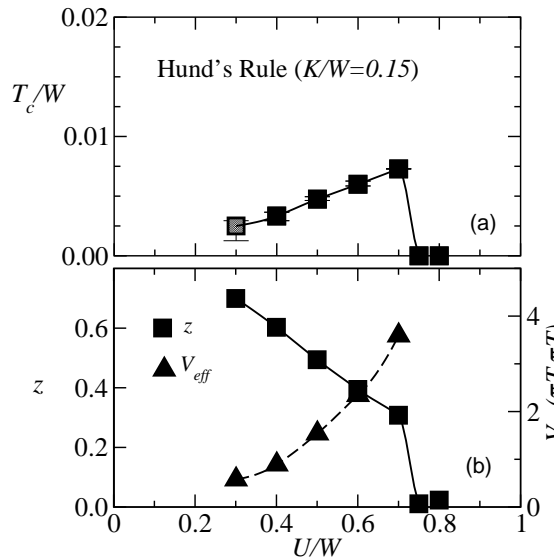}}}
\caption{(a) Superconducting transition temperature $T_c$ vs Coulomb 
repulsion $U$ with Hund's rule coupling. 
For the Hund's rule (HR) coupled system, $T_c$ monotonically increases
with $U$ showing local pairing mechanism. Gray symbol is an extrapolated
estimate from the superconducting kernel $\chi_0\Gamma$.
(b) Quasi-particle renormalization factor $z$
(left axis) and effective pairing interaction $V_{\rm eff}$ at lowest
Matsubara frequency(right axis).
$z$ decreases as $U$ approaches $U_c$.
The product $zV_{\rm eff}$ for the HR system increased
as $U\rightarrow U_c$, consistent with the increasing $T_c$.
$W$ is the non-interacting bandwidth.
}
\label{fig:tchr}\end{figure}

\begin{figure}[bt]
\rotatebox{0}{\resizebox{3.0in}{!}{
\includegraphics{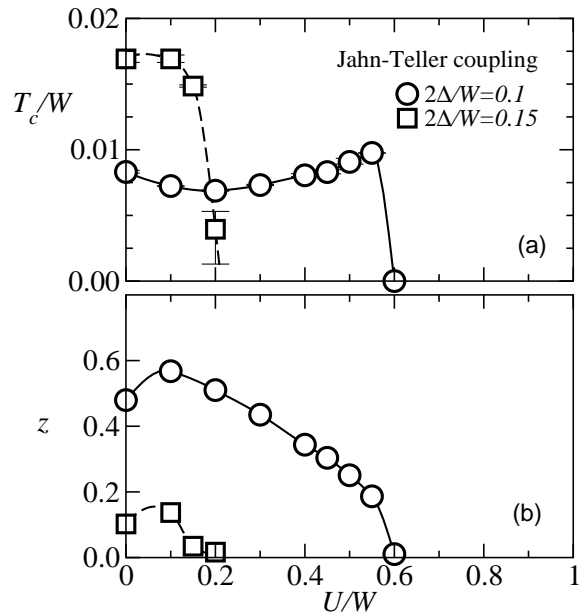}}}
\caption{(a) Superconducting transition temperature $T_c$ vs Coulomb
repulsion $U$ with Jahn-Teller electron-phonon coupling.
The Jahn-Teller (JT) phonon coupling 
produces superconductivity with a cross-over from the
conventional superconductivity (small $U$) to the local pairing 
regime (large $U$) for $2\Delta/W=0.1$. For a larger electron-phonon
coupling ($2\Delta/W=0.15$), the Mott insulator transition happened
at smaller $U$. (b) Renormalization factor $z$ as in Fig.~\ref{fig:tchr}.
}
\label{fig:tcjt}\end{figure}

Fig.~\ref{fig:tchr}(a) shows transition temperatures
$T_c$ for the HR coupling as a function of the Coulomb repulsion $U$ 
where the superconductivity becomes stronger with $U$.
The HR superconductivity emerged only in the spin-triplet channel.
$T_c$ {\it monotonically increased} as $U$ approached
the critical value $U_c/W\approx 0.75$ for $K/W=0.15$,
a strong indication of 
the local pairing. $T_c/W$ less than 0.002 (gray symbol) has been estimated
by extrapolating eigenvalues of the superconducting kernel 
$\sqrt{\chi_0}\Gamma\sqrt{\chi_0}$. 
The HR-coupled superconductivity is strong only in the local pairing 
regime, since at small $U$ there exists no {\em external} low-energy 
pairing medium in the electronic HR coupling.

$V_{\rm eff}(\pi T,\pi T)$ at the lowest Matsubara frequency
at $T/W=1/120$ (triangles in Fig.~\ref{fig:tchr}(b), to the 
right-hand-side scale)
show a growing local pairing attraction as $z\rightarrow 0$~\cite{veff}.
The product $zV_{\rm eff}(\pi T,\pi T)\,(\propto\chi_0 
\Gamma)$ increases as $U\rightarrow U_c$,
consistent with the increasing $T_c$.
We emphasize that the local pairing
becomes apparent even at moderately correlated regime of $z\sim 0.5$, 
although the heuristic argument in section~\ref{sec:pairing}
was given for $z\rightarrow 0$ for the sake of clarity.

We can make some predictions based on general properties of the
local spin-triplet pairing. (1) The exchange anti-symmetry is in the orbital
space, in contrast to the $p$-wave spin-triplet theory~\cite{balian}.
Due to the discreteness of orbitals, the quasi-particle excitation is
gapped. The spatial part has the $s$-wave symmetry. (2) For large
orbital degeneracy ($N_d>2$), coupling between different pairs
of orbital indices will likely result in multiple gaps. 
(3) The local pairing is strong for high electron/hole 
dopings~\cite{han_sc}.
For small dopings, ferromagnetic fluctuation will dominate the local
triplet pairing. (4) As can be inferred from Fig~\ref{fig:dia},
strong ferromagnetic order suppress the pair-exchange. Therefore,
if the pairing and ferromagnetic phases ever co-exist, the 
superconducting phase should be near the ferromagnetic phase boundary.

Jahn-Teller phonon coupling (in Fig.~\ref{fig:tcjt})
shows the similar behavior of the
local pairing as the HR pairing. 
In contrast to the HR coupling case, $T_c$ is non-zero at $U=0$
with the pairing mediated by the {\em external} coupling medium of phonons.
The results for $2\Delta/W=0.1$
shows that the phonon pairing-medium
provides the attractive pairing for the small $U$ and as
$U$ is increased further the superconductivity makes a cross-over
to the local pairing regime. 
At large electron-phonon coupling ($2\Delta/W=0.15$) the metallic
region shrunk and showed no signs of the cross-over.
Given the same coupling strength (comparing $K$ and $|K_{\rm ph}|=
2\Delta$), the HR system has weaker superconductivity
than the el-ph systems. 

One significant difference between this
model of doubly degenerate orbitals (commonly called as
$e\times E$ model) and the C$_{60}$ model ($t\times H$ 
model~\cite{han_sc}) is that the $e\times E$ model produces
stronger enhancement of $T_c$ in the local pairing regime. 
Authors in Ref.~\cite{han_sc} found
no signs for the up-turn of $T_c$ near the Mott transition for
any studied values of $\Delta$ in the C$_{60}$-model, although 
the enhancement of $T_c$
over the conventional theories was evident. It is probably due to
the easier low spin formation for even fillings, since at odd
number of fillings (as 3 in the C$_{60}$ model) unpaired spins
tend to fluctuate via the intersite hopping and make the low
spin configuration less effective.

Since the HR and JT couplings induce superconductivities separately
but in different symmetries, they are expected to complete
when both are present. For instance, by turning up the JT coupling
constant at a fixed finite HR coupling, one can suppress the spin-triplet
pairing and eventually crosses over to a spin-singlet pairing.
It can be also anticipated from the
mapping of the JT coupling to a negative HR hamiltonian
in the anti-adiabatic limit. 
Fullerene systems are known to have sizable strengths for both
interactions, although the JT interaction is more dominant for 
alkali-doped region.
A recent density functional calculation study~\cite{luders} 
has suggested that the fullerene in the hole-doped region may
have stronger Hund's rule interaction, with possible
applications to cation-doped fullerenes~\cite{holedoped}.

Finally, we make an observation of an interesting property of the
local interaction Eq.~(\ref{eq:hjt}). 
If we swap the spin and orbital indices
in the JT coupling, Eq.~(\ref{eq:hjt}), new boson fields 
$\tilde{\varphi}_\nu$ couple to electronic spin fluctuations, 
{\it i.e.}, the coupling terms become
$\sum_m[\tilde{\varphi}_1(n_{m\uparrow}-n_{m\downarrow})
  +\tilde{\varphi}_2(d^\dagger_{m\uparrow}d_{m\downarrow}+{\rm h.c.})]$.
Now $\tilde{\varphi}_\nu$ plays the role of {\it external} local
spin-fluctuations which mediate the electron pairing.
All other terms in the hamiltonian are unchanged. The ground state
now belongs in the spin-triplet and orbital-singlet space.
The pair susceptibility for the singlet channel 
maps to that of the triplet-channel
Eq.~(\ref{eq:chiT}) and all results from JT coupling automatically
hold for the triplet superconductivity. This model provides a robust
example for the local spin-fluctuation superconductivity.

\section{Conclusion}

We have demonstrated that the local pairing mechanism in
multi-band systems on general grounds. In particular, the existence of 
Hund's rule induced spin-triplet $s$-wave superconductivity is shown. 
The self-generated local multiplet fluctuations in the strong-$U$
limit provide the pairing medium even with
a purely electronic interaction. This general idea should work
for other extensions of the model.
Further studies on the superconductor-ferromagnet phase diagram 
would clarify the relevance of the current model to known experimental
systems.
Fluctuation effects beyond the mean field approximation and
singlet-triplet competition are left for future research. 

\begin{acknowledgments}
We thank O. Gunnarsson, E. Tosatti, M. Fabrizio and V. H. Crespi for 
helpful discussions. We acknowledge support from the National Science 
Foundation DMR-9876232 and Max-Planck forschungspreis.
\end{acknowledgments}

\appendix

\section{Atomic level scheme of Hund's rule coupling}
\label{app1}
For 2 electrons on site, there are 6 possible configurations
\begin{equation}
\left\{d^\dagger_{1\up}d^\dagger_{2\up},\,d^\dagger_{1\dn}d^\dagger_{2\dn},\,
d^\dagger_{1\up}d^\dagger_{2\dn},d^\dagger_{2\up}d^\dagger_{1\dn},
d^\dagger_{1\up}d^\dagger_{1\dn},d^\dagger_{2\up}d^\dagger_{2\dn}
\right\}|vac\rangle.
\end{equation}
These states are degenerate for $H_{\rm mult}=0$.
$H_{\rm mult}=H_{\rm HR}$ of Eq.~(\ref{eq:hhr}) can be rewritten
in the above basis set as
\begin{equation}
\left(\begin{array}{cccccc}
-2K & 0 & 0 & 0 & 0 & 0 \\
0 & -2K & 0 & 0 & 0 & 0 \\
0 & 0 & -K & K & 0 & 0 \\
0 & 0 & K & -K & 0 & 0 \\
0 & 0 & 0 & 0 & K & K \\
0 & 0 & 0 & 0 & K & K \end{array}\right),
\end{equation}
the solutions of which become
\begin{equation}
\begin{array}{ll}
E=-2K, & \left\{\begin{array}{r}
             d^\dagger_{1\up}d^\dagger_{2\up}\\
             1/\sqrt{2}(d^\dagger_{1\up}d^\dagger_{2\dn}
             -d^\dagger_{2\up}d^\dagger_{1\dn})\\
             d^\dagger_{1\dn}d^\dagger_{2\dn}\end{array}\right\}|vac\rangle \\
E=0, & \left\{\begin{array}{r}
             1/\sqrt2(d^\dagger_{1\up}d^\dagger_{2\dn}
             +d^\dagger_{2\up}d^\dagger_{1\dn})\\
             1/\sqrt2(d^\dagger_{1\up}d^\dagger_{1\dn}
             -d^\dagger_{2\up}d^\dagger_{2\dn})\end{array}\right\}|vac\rangle \\
E=2K, & 1/\sqrt2(d^\dagger_{1\up}d^\dagger_{1\dn}
             +d^\dagger_{2\up}d^\dagger_{2\dn})|vac\rangle.
\end{array}
\end{equation}
As shown in Fig.~\ref{fig:scheme}, the ground states are $S=1$ spin-triplet
and the excited states are $S=0$ spin-singlet with orbitally mixed states.
The above ground multiplet states are written as $|T_1),\,|T_0)$ and 
$|T_{\bar1})$ for the triplet states ($T$) of $S_z=1,0,-1$, respectively.
Similarly, we write the above excited states as $|S_a),\,|S_b)$ and
$|S_c)$, respectively, for spin-singlets ($S$).
With one electron on site, the Hund's rule coupling does not lift any
degeneracy since it is a multi-electron interaction. The same is true
with 3 electrons, since one can view this in terms of the hole interaction.

\section{Perturbation theory for pairing correlation energy}
\label{app2}

The perturbation is defined as an expansion with respect to the
intersite hopping around the interacting Hamiltonian in
Eq.~(\ref{eq:heff}),
\begin{eqnarray}
  H_0 & = & UN(N-1)/2+H_{\rm mult}\\
  V & = & -t_q \sum_{m\sigma}(c^\dagger_{m\sigma}d_{m\sigma}+h.c.).
\end{eqnarray}
First, let us develop a general perturbation theory up to the 4-th 
order. The Heisenberg equation reads
\begin{equation}
 H|\psi\rangle = (H_0+V)|\psi\rangle = E|\psi\rangle,
\label{app:eigen}
\end{equation}
where the energy eigenvalue and wavefunction are
expanded in terms of the perturbation $V$ as
\begin{eqnarray}
 E & = & E_0 + E_1 + E_2 + E_3 + E_4,\\
 |\psi\rangle & = & |\psi_0\rangle + |\psi_1\rangle + |\psi_2\rangle 
              + |\psi_3\rangle + |\psi_4\rangle.
\end{eqnarray}
The 0-th order terms give
\begin{equation}
 H_0|\psi_0\rangle = E_0|\psi_0\rangle.
\end{equation}
The $n$-st order equation becomes
\begin{equation}
 H_0|\psi_n\rangle + V|\psi_{n-1}\rangle = \sum_{m=0}^{n}E_m|\psi_{n-m}\rangle.
\label{app:expand}
\end{equation}
Projecting the unperturbed bra-vector $\langle\psi_0|$ to 
Eq.~(\ref{app:expand}) for $n=1$, 
one gets the first order correction to the energy
\begin{equation}
E_1=\langle\psi_0|V|\psi_0\rangle.
\end{equation}
Introducing an operator $P_0$ which projects out the unperturbed wavefunction
$|\psi_0\rangle$, one can express the wavefunction to the 1st order as,
\begin{equation}
 |\psi_1\rangle = {1\over E_0-H_0}P_0V|\psi_0\rangle.
\label{app:psi1}
\end{equation}
The higher order contributions are computed by induction. The useful
expressions up to the 4th order are
\begin{eqnarray}
|\psi_2\rangle & =  & {1\over E_0-H_0}P_0(V-E_1)|\psi_1\rangle\label{app:psi2}\\
 E_2 & = & \langle\psi_0|V|\psi_1\rangle = 
           \langle\psi_1|(E_0-H_0)|\psi_1\rangle \label{app:e2}\\
 E_4 & = & \langle\psi_2|(E_0-H_0)|\psi_2\rangle-E_2\langle\psi_1
      |\psi_1\rangle,\label{app:e4}
\end{eqnarray}
where we have used a simplification $E_1=0$ which is the case with our models.

The unperturbed energies with the Jahn-Teller interaction, 
Eq.~(\ref{eq:hjt}), at the total occupation numbers $N_d,N_d+1,N_d+2$ are
\begin{eqnarray}
E(N_d+2) & = & U-8\Delta-{4t_q^2\over U}-{14t_q^4\over \Delta U^2} \\
E(N_d+1) & = & U-8\Delta-{4t_q^2\over U}-{7t_q^4\over 2\Delta U^2} \\
E(N_d) & = & U-8\Delta-{4t_q^2\over U},
\end{eqnarray}
where $\Delta=-8g^2/\omega_{\rm ph}$ is the local Jahn-Teller coupling energy.
Corresponding wavefunctions of the leading order are
\begin{eqnarray}
|\psi(N_d+2)\rangle & = & |S_c)\otimes|s_c) \\
|\psi(N_d+1)\rangle & = & |S_c)\otimes c^\dagger_{m\sigma}|vac) \\
|\psi(N_d)\rangle & = & |S_c)\otimes |vac),
\end{eqnarray}
with notations given in Appendix~\ref{app1}.

\section{Hubbard-Stratonovich decoupling for Hund's rule coupling}
\label{app:hs}

The discrete Hubbard-Stratonovich transformation~\cite{hirsch}(DHST) is given
by
\begin{equation}
e^{-\Delta\tau Un_\alpha n_\beta}=e^{-{\Delta\tau U\over 2}
         (n_\alpha+n_\beta)}{1\over 2}
    \sum_{\sigma}e^{\lambda\sigma(n_\alpha-n_\beta)},
\label{dhst}
\end{equation}
for fermion number operator $n_\alpha$ for a quantum state $\alpha$ and 
$\cosh\lambda=\exp(\Delta\tau U/2)$. The continuous Hubbard-Stratonovich 
transformation~\cite{bss}(CHST) applies to a general operator $\hat{A}$
in a gaussian integral
\begin{equation}
e^{\hat{A}^2}=\int dx\,\exp(-\pi x^2+\sqrt{\pi}\hat{A}x).
\label{chst}
\end{equation}
For doubly degenerate orbital systems, the interacting Hamiltonian 
Eq.~(\ref{eq:hhr}) can be rewritten as
\begin{eqnarray}
H_{\rm int} & = & {U\over 2}\left[N_+ + N_-\right]^2 \label{app:hhr} \\
      & + & K\left[\sum_{m}n_{m\uparrow}n_{m\downarrow}
                 \!-\!\!\!\sum_{m\neq m'}\!\!n_{m\uparrow}n_{m'\downarrow}
                 \!-2\!\!\!\!\sum_{\sigma,m<m'}\!\!\!\!n_{m\sigma}
		 n_{m'\sigma}\right] \nonumber \\
           & + & K\!\!\sum_{m\neq m'}\!\!\left[
                 d^\dagger_{m\uparrow}d^\dagger_{m\downarrow}
                 d_{m'\downarrow}d_{m'\uparrow}
                +d^\dagger_{m\uparrow}d^\dagger_{m'\downarrow}
                 d_{m\downarrow}d_{m'\uparrow}\right], \nonumber
\end{eqnarray}
where $N_+=\sum_m n_{m\uparrow}$ and $N_-=\sum_m n_{m\downarrow}$.
Previously, each terms in the first two lines have been
decoupled by the DHST. $N_+^2,N_-^2,N_+N_-$ are expanded
into $n_{m\sigma}n_{m'\sigma'}$ terms before the DHST Eq.~(\ref{dhst})
is applied.
The terms on the third line have been decoupled by
\begin{equation}
 e^{-\Delta\tau K c^\dagger_\alpha c^\dagger_\beta c_\delta c_\gamma}
   ={1\over 2}
   \sum_{\sigma}e^{\sqrt{\Delta\tau  K}\sigma(c^\dagger_\alpha
    c_\gamma-c^\dagger_\beta c_\delta)},
\end{equation}
where $K>0$ and no pairs of the indices from $\{\alpha,\beta,\gamma,\delta\}$ 
are the same. When the QMC sampling is incomplete, all the symmetry
inherent in the original hamiltonian is not fully recovered, which
is speculated to be a source of the sign-problem.
This problem can be resolved by a different decoupling scheme as follows.
We rewrite the first line as
\begin{equation}
  {U\over 2}(N_++N_-)^2=U(N_+^2+N_-^2)-{U\over 2}S_z^2,
\label{app:line1}
\end{equation}
with $S_z=N_+-N_-$.
For the orbital degeneracy $N_d=2$, the first two terms in the second 
line in Eq.~(\ref{app:hhr}) can be written as
\begin{eqnarray}
 & & K\left(n_{1\up}n_{1\dn}+n_{2\up}n_{2\dn}-n_{1\up}n_{2\dn}
    -n_{2\up}n_{1\dn}\right) \nonumber \\
 & = & -{K\over 2}\left(n_{1\up}-n_{2\up}-n_{1\dn}+n_{2\dn}\right)^2
  \nonumber \\
 & & +{K\over 2}(N_++N_-)-K(n_{1\up}n_{2\up}+n_{2\dn}n_{1\dn}),
\label{app:line2}
\end{eqnarray}
where the first term with squared term is decoupled with the CHST
of a single auxiliary field. This procedure preserves the symmetry
between orbital 1 and 2. The third line in Eq.~(\ref{app:hhr}) has
been the source of the sign-problem~\cite{han_DMFT}. 
We express this by completing square as
\begin{eqnarray}
& &K(c^\dagger_{1\up}c^\dagger_{1\dn}c_{2\dn}c_{2\up}+
    c^\dagger_{2\up}c^\dagger_{2\dn}c_{1\dn}c_{1\up}+\cdots ) \nonumber \\
& = & -{K\over 2}(c^\dagger_{1\up}c_{2\up}+c^\dagger_{2\up}c_{1\up}
           -c^\dagger_{1\dn}c_{2\dn}-c^\dagger_{2\dn}c_{1\dn})^2 \nonumber \\
& & +{K\over 2}(N_++N_-)-K(n_{1\up}n_{2\up}+n_{2\dn}n_{1\dn}).
\label{app:line3}
\end{eqnarray}
Similarly to Eq.~(\ref{app:line2}), the squared term preserves 
the orbital symmetry with a negative coefficient ($K>0$). With 
$-2(n_{1\up}n_{2\up}+n_{2\dn}n_{1\dn})=-(N_+^2+N_-^2)+(N_++N_-)$
and adding Eqs. (\ref{app:line1}-\ref{app:line3}), we finally have
\begin{eqnarray}
H_{\rm int}\!\! & =\!\! & 3K(N_++N_-)+(U-2K)(N_+^2+N_-^2)\nonumber \\
  & & -{U\over 2}S_z^2-{K\over 2}\left(n_{1\up}-n_{2\up}-n_{1\dn}
      +n_{2\dn}\right)^2 \nonumber \\
  & & -{K\over 2}(c^\dagger_{1\up}c_{2\up}+c^\dagger_{2\up}c_{1\up}
           -c^\dagger_{1\dn}c_{2\dn}-c^\dagger_{2\dn}c_{1\dn})^2,
\end{eqnarray}
where the $N_\pm$ in the first line are expanded into 
$n_{m\sigma}n_{m'\sigma'}$ and decoupled by the DHST, Eq.~(\ref{dhst}). 
Terms in
the second and third lines are decoupled by the CHST, Eq.~(\ref{chst}). 
With the old DHST scheme for all interaction terms, there are 6 
auxiliary fields for $n_\alpha n_\beta$ pairs and 4 for 
$c^\dagger_\alpha c^\dagger_\beta
c_\gamma c_\delta$ terms, 
10 in total per each time slice. The improved scheme has 2 discrete
fields for $n_{1\up}n_{2\up}$, $n_{1\dn}n_{2\dn}$ in $N_{\pm}^2$ and
3 continuous fields per time slice.
This procedure can be extended to higher degeneracy $N>2$.
Although the arrangement in Eq.~(\ref{app:line2}) is special
for $N_d=2$, they can be decoupled by the DHST
without causing the sign-problem.
Terms like in Eq.~(\ref{app:line3}) are
problematic for the sign-problems in the impurity problem.
Eq.~(\ref{app:line3}) can be readily extended to higher degeneracy
and the above procedure will likely remove the sign-problem
in the Hund's rule coupling.


\begin{thebibliography}{*}
\bibitem{veff} $V_{\rm eff}$ is the effective interaction in terms of
	non-interacting basis instead of between quasi-particles
	in $E_p$, Eq.~(\ref{eq:epair}).
\end{thebibliography}
\end{document}